# Structural routes to stabilize superconducting $La_3Ni_2O_7$ at ambient pressure

Luke C. Rhodes[1,*] and Peter Wahl[1,2]

[1]*SUPA, School of Physics and Astronomy, University of St Andrews, North Haugh, St Andrews, KY16 9SS, United Kingdom*
[2]*Physikalisches Institut, Universität Bonn, Nussallee 12, 53115 Bonn, Germany*

The bilayer perovskite $La_3Ni_2O_7$ has recently been found to enter a superconducting state under hydrostatic pressure at temperatures as high as 80 K. The onset of superconductivity is observed concurrent with a structural transition which suggests that superconductivity is inherently related to this specific structure. Here we perform density functional theory based structural relaxation calculations and identify several promising routes to stabilize the crystal structure which hosts the superconducting state at ambient pressure. We find that the structural transition is controlled almost entirely by a reduction of the *b*-axis lattice constant, which suggests that uniaxial compression along the [010] direction or in-plane biaxial compression are sufficient as tuning parameters to control this transition. Furthermore, we show that increasing the size of the *A*-site cations can also induce the structural transitions via chemical pressure and identify $Ac_3Ni_2O_7$ and Ba-doped $La_3Ni_2O_7$ as potential candidates for a high temperature superconducting nickelate at ambient pressure.

## I. INTRODUCTION

The recent discovery of an 80 K superconducting state in $La_3Ni_2O_7$ under pressure has significantly expanded the ability to study high-temperature superconductivity in perovskite systems [1]. Unlike the infinite layer nickelates, which become superconducting at temperatures below 10 K [2,3], $La_3Ni_2O_7$ exists in bulk single crystals, enabling similar systematic studies to identify how superconductivity evolves under experimental parameters such as doping, strain, and magnetic field [4–6].

The structural and electronic similarity of $La_3Ni_2O_7$ to the cuprate and ruthenate superconductors is also highly suggestive that $La_3Ni_2O_7$ could host an unconventional superconducting mechanism. This has motivated theoretical efforts to understand the nature of superconductivity in this system using a strong coupling bilayer Hubbard model [7–20] and the random phase approximation (RPA) [21–23], as well as exploring the effect of correlations on the electronic structure [24–28]. However, unique to this system, the stability of the superconducting state is very sensitive to the underlying crystal structure. Specifically, small tilting of $NiO_6$ octahedra appears to suppress superconductivity, which emerges only at high pressure once the tilting is lifted, concomitant with a structural transition from a crystal structure with *Amam* to one with *Fmmm* space group [1]. There is thus an important open question regarding the origin of this structural transition and its role in stabilizing superconductivity. We demonstrate here routes to stabilize the *Fmmm* structure, which hosts the high-temperature superconducting state at ambient pressure.

Density functional theory (DFT) provides an ideal method to answer this question; despite not capturing the strong correlation effects that are thought to be required to induce superconductivity [1], structural relaxations are usually (although not always [29–32]) governed by chemical bonding and electronic states on much larger energy scales. Consequently, DFT is known to correctly identify the structural motifs of similar Ruddlesden-Popper Perovskite structures, e.g., the octahedral rotations in the surface layer of $Sr_2RuO_4$ [33,34] and the pressure induced structural transition in $Ca_2RuO_4$ [35].

Here, we perform structural relaxations using DFT to explore different routes to control the crystal structure. As tuning parameters, we consider uniaxial and biaxial strain as well as hydrostatic and chemical pressure. Our systematic study allows us to elucidate the mechanism behind the structural transition and how it can be influenced and controlled. We find that the transition is controlled almost entirely by a reduction of the *b* lattice parameter, suggesting that uniaxial (or biaxial) strain are alternative routes to stabilize the same crystal structure that hosts the superconducting ground state at high pressure. Alternatively, through chemical pressure, by replacing the La ion with a larger cation, either Ba or Ac, the *Fmmm* structure can be stabilized at ambient pressure.

## II. METHODS

We use the Vienna *ab initio* simulation package (VASP) [36–38] to perform our DFT simulations, using the PBE version of the generalized gradient approximation as the exchange correlation functional [39] and projected augmented wave (PAW) pseudopotentials [40]. As a starting point, we took the experimental crystal structure from Sun *et al.* [1]

---

*Corresponding author: lcr23@st-andrews.ac.uk

and relaxed both the lattice constants and internal degrees of freedom. To simulate the effect of uniaxial and biaxial strain as well as hydrostatic pressure, we directly modify the lattice constants of this structure by a linear scaling factor, assuming a negligible Poisson ratio, and then relax the internal degrees of freedom only. We have verified that the inclusion of a nonzero Poisson ratio does not qualitatively change the conclusions presented here. While VASP allows one to include hydrostatic pressure explicitly, as has been previously performed on $La_3Ni_2O_7$ [21,41,42], by setting the lattice constants by hand, we can directly compare the effect of hydrostatic pressure simulations with that of uniaxial and biaxial strain. A wave function cutoff (ENCUT) of 600 eV was used throughout, as well as a $k$ grid of $8 \times 8 \times 8$. All structural relaxations were performed using an energy convergence threshold of $10^{-6}$ eV and a force threshold of $10^{-4}$ eV Å$^{-1}$.

## III. RESULTS

To begin our analysis, we first require the DFT-relaxed ground state crystal structure. We thus have performed a full relaxation of the bulk crystal structure, including both internal and lattice degrees of freedom, from the experimentally determined lattice parameters of $La_3Ni_2O_7$ from Sun *et al.* [1], and find that the $a$, $b$, $c$ lattice constants relax to $a = 5.414$ Å, $b = 5.5218$ Å, and $c = 20.2517$ Å. The $a$ and $c$ lattice constants are within 0.4% of the measured values at 1.6 GPa, whereas the $b$ value is 1.3% larger. Additionally, we find a Ni-O-Ni bond angle off the high symmetry $c$ axis ($\theta_c$) of 8.1°, compared to the 6.25° inferred from x-ray crystallography [1]. We find that this error is reduced when introducing mean-field electron correlation effects via the DFT+$U$ extension; see Appendix A and Table I. However, the qualitative trends are not dependent on such a term. For this reason the results presented in the main text are performed without the inclusion of a Hubbard $U$ term.

At ambient pressure, the crystal structure is in the *Amam* space group, which is characterized by a $\theta_c \neq 0°$ [Fig. 1(a)]. However, under hydrostatic pressure the *Fmmm* space group is stabilized [characterized by $\theta_c = 0$, shown in Fig. 1(b)]. By reducing the $a$, $b$, and $c$ lattice constants linearly, we simulate the effect of pressure and capture a smooth transition of the crystal structure from the *Amam* to *Fmmm* space group as shown in Fig. 1(c). The transition is observed once the $a$, $b$, and $c$ lattice constants have been reduced by 3.25%, or equivalently a 10.1% volume reduction. By calculating the stress tensor, we infer that this corresponds to an external pressure of 20 GPa. This finding is qualitatively consistent with the experimental measurement of Sun *et al.* [1]; however, they observe the transition for a volume reduction of 5%, and at 14 GPa about a factor of 2 smaller than where it occurs in these DFT calculations. Again, this error can be quantitatively reduced by the inclusion of a Hubbard $U$ term [1,41,42], as discussed in Appendix A, or with antiferromagnetic order [24]. However, the transition is still observed for a nonmagnetic calculation without the inclusion of a Hubbard $U$ term, confirming that the structural transition is not primarily driven by electronic correlations.

The impact of the structural transition on the electronic structure is subtle. The low energy electronic structure near the Fermi energy is dominated by Ni $e_g$ bands derived from the Ni $d_{x^2-y^2}$ and $d_{z^2}$ orbitals. Consequently, spin-orbit coupling will only have a negligible effect on the bands close to the Fermi energy. The band structure at low pressure is shown in Fig. 1(d). It is predominantly 2D in character, with some $k_z$ dispersion along the $\Gamma$-$Z$ direction, which is evidence for non-negligible interlayer coupling. The small rotation of the oxygen atoms around the Ni atom mixes $d_{z^2}$ and $d_{x^2-y^2}$ orbital character and results in opening of small hybridization gaps at crossings of the $d_{x^2-y^2}$- and $d_{z^2}$-derived bands. On entering the *Fmmm* phase, with the straightening of the $c$-axis Ni-O-Ni bond, the mixing between the $d_{z^2}$- and $d_{x^2-y^2}$-derived bands is suppressed, as shown in Fig. 1(e), and the hybridization gap closes. This mechanism for opening hybridization gaps is a general argument, based on simple linear combination of atomic orbitals and their overlaps. It may be relevant for the onset of the superconducting state, as discussed in Ref. [43].

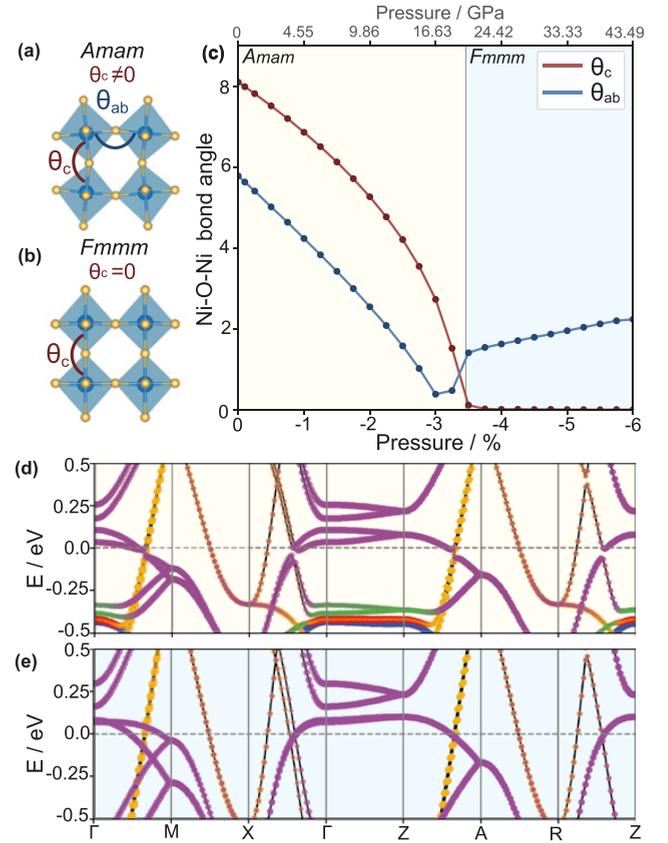

FIG. 1. (a), (b) Crystal structure of $La_3Ni_2O_7$ with Ni (blue), oxygen (yellow) in space group *Amam* (a) and *Fmmm* (b). (c) Comparison of the Ni-O-Ni bond angle parallel to the $c$ axis (red, $\theta_c$) and $ab$ plane (blue, $\theta_{ab}$) as a function of hydrostatic pressure (percentage reduction of the $a$, $b$, $c$ lattice constants) from DFT relaxation simulations. The in-plane Ni-O-Ni bond angle evolves from concave to convex as pressure is increased (lattice constants reduced) with an inflection point at 3%. (d) Low energy band structure of $La_3Ni_2O_7$ with *Amam* symmetry from the relaxed structure at 0% pressure and (e) the *Fmmm* band structure from the relaxed structure at 4% pressure. corresponding to approximately 24 GPa hydrostatic pressure. (e) The bands are plotted along the high symmetry **k** path $\Gamma = (0, 0, 0)$, $M = (\pi, \pi, 0)$, $X = (\pi, 0, 0)$, $Z = (0, 0, \pi)$, $A = (\pi, \pi, \pi)$, and $R = (\pi, 0, \pi)$. Purple and yellow coloring correspond to $d_{z^2}$ and $d_{x^2-y^2}$ dominant orbital character, respectively; other colors represent the $t_{2g}$ orbitals.

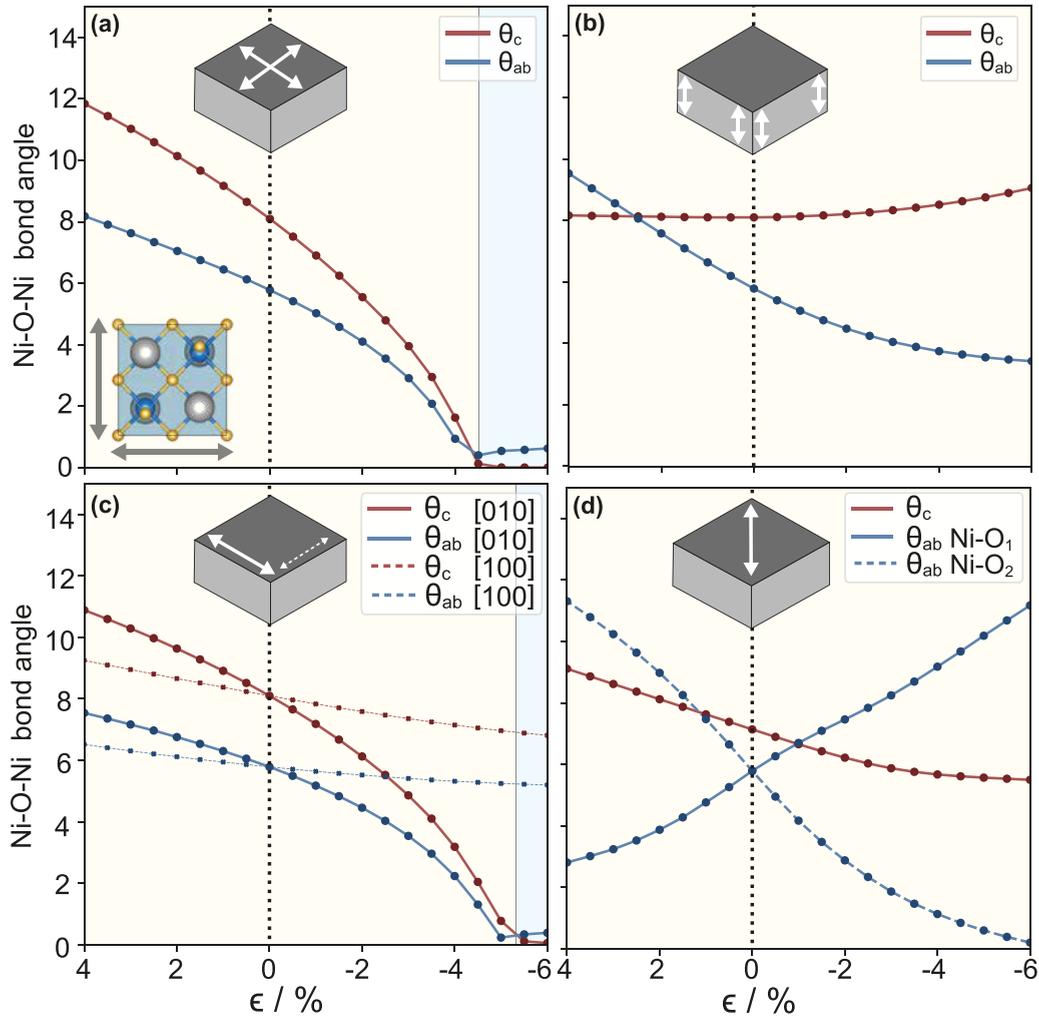

FIG. 2. Effect of biaxial and uniaxial strain ($\epsilon$) on the structural transition of La$_3$Ni$_2$O$_7$. Evolution of the in-plane and out-of-plane Ni-O-Ni bond angles as a function of (a) biaxial strain along [100] and [010]. (b) Uniaxial strain along *c*. (c) Uniaxial strain along the [010] (solid lines) and [100] (dashed lines) directions, 45° to the Ni-Ni direction. (d) Uniaxial strain along the [110] direction (Ni-Ni direction). Note that this breaks the symmetry of the Ni-O-Ni bonds along the [110] and [1−10] directions so both are plotted. The light blue color indicates the *Fmmm* lattice symmetry. An inset of the strain with relation to the crystal structure has been included in (a), with the La, Ni, and O atoms colored as gray, blue, and yellow, respectively.

Recent ARPES measurements [44] place the flat band around the Γ point slightly below the Fermi level, which can be captured within DFT+*U* [1,42,44]. Although this may be important for the stabilization of superconductivity, it does not appear to be a prerequisite for the structural transition.

To identify ways to stabilize the crystal structure that hosts the superconducting state, we explore biaxial and uniaxial strain (Fig. 2). We vary the lattice constants in the direction of strain $\epsilon = \frac{\delta a}{a}$ linearly, assuming a Poisson ratio of zero, but note that using other values for the Poisson ratio does not change the qualitative behavior. We find that by decreasing the *a* and *b* lattice constants while keeping the *c*-lattice constant fixed (biaxial compression), we again stabilize the *Fmmm* symmetry at $\epsilon > 4.5\%$ [Fig. 2(a)]. Straining the sample by making the lattice constants larger (biaxial strain) only increases the Ni-O-Ni bond angle.

For out-of-plane uniaxial strain, when we only change the *c* lattice constant, the structure does not evolve to a structure compatible with the *Fmmm* space group, for either expansion or compression up to ±6% [Fig. 2(b)].

With in-plane uniaxial strain, there is a choice of axis in which the strain can be applied: either [100] or [010], as shown in Fig. 2(c), or [110] as shown in Fig. 2(d). Uniaxial strain along the [100] direction is insufficient to induce the transition of the crystal structure to the *Fmmm* space group within ±6% strain (defined by $\theta_c = 0$). However, applying strain along the [010] direction directly induces the transition to a structure with the *Fmmm* space group at a compressive strain close to 5%. While this is a very large value, we expect the transition in reality to occur for a smaller strain, due to the factor of 2 difference between the strain at which the transition occurs in our simulations and the experimental results under hydrostatic pressure. Repeating this calculation within the DFT+*U* framework reduces this transition to 4% as discussed in Appendix A. Applying the strain along the

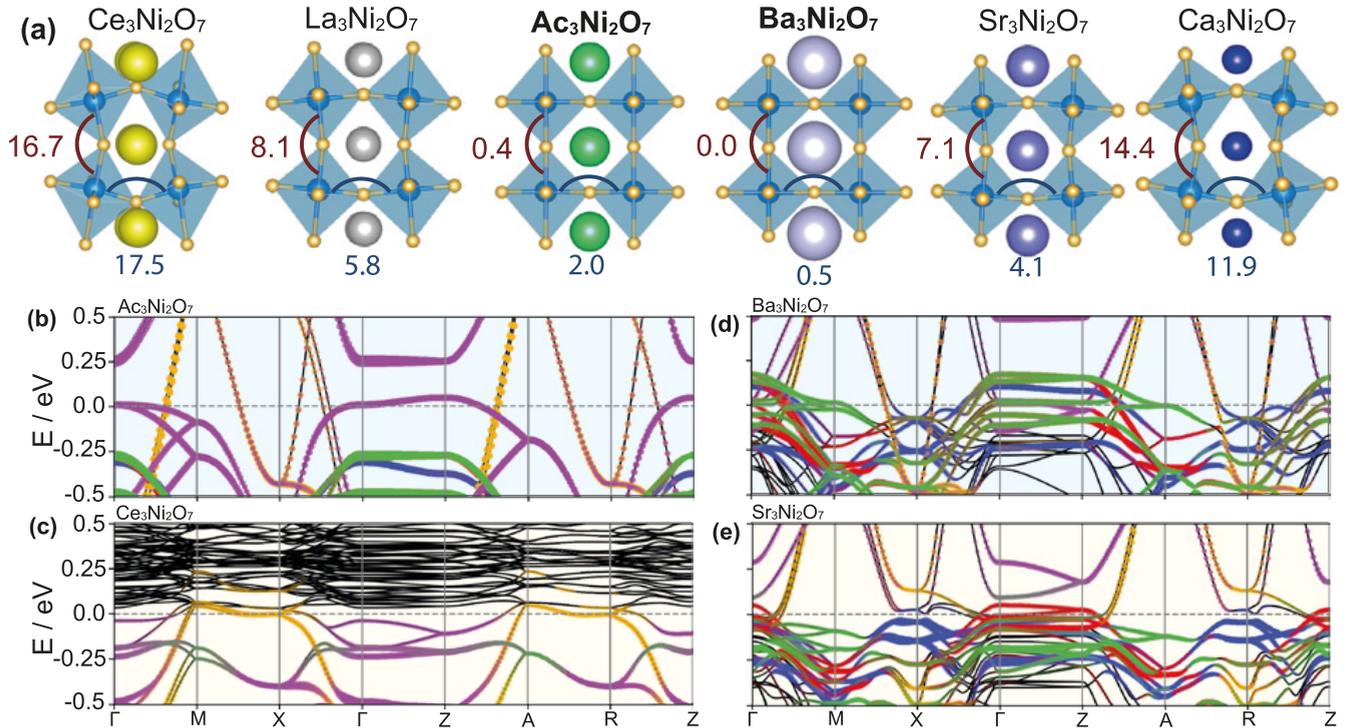

FIG. 3. Influence of the $A$-site cation on the structure of $A_3Ni_2O_7$. (a) Relaxed structure of $A_3Ni_2O_7$ for six different $A$-site cations, $A =$ Ce, La, Ac, Ba, Sr, Ca. The out-of-plane and in-plane Ni-O-Ni bond angles ($\theta_c$ and $\theta_{ab}$, respectively) are also shown. The corresponding band structures within 500 meV of the Fermi level are shown in (b)–(e) for (b) $Ac_3Ni_2O_7$, (c) $Ce_3Ni_2O_7$, (d) $Ba_3Ni_2O_7$, and (e) $Sr_3Ni_2O_7$. Red, green, and blue lines correspond to bands with Ni $3d_{xz}$, $3d_{yz}$, and $3d_{xy}$ orbital character, whereas yellow and purple correspond to the $3d_{x^2-y^2}$ and $3d_{z^2}$ orbitals, respectively. The black bands in (c) are $4f$-derived bands from the Ce atom.

[110] direction is also not sufficient to induce the structural transition [Fig. 2(d)], as it only modifies $\theta_c$ by 1.25° between $-4\%$ and $+6\%$ uniaxial strain. We therefore find that the structural transition is controlled by the compression of the $b$-axis lattice constant. This intuitively can be explained as reducing the $NiO_6$ octahedral tilt in the $Amam$ space group which is also along the $b$-axis direction.

The final tuning parameter that we investigate here is chemical pressure. Like in many metallic Ruddlesden-Popper materials with formula $A_{n+1}B_nO_{3n+1}$, the low energy electronic properties are goverened almost entirely by the $B$-site valence shell, which in this case are the Ni $3d$ orbitals. Changing the $A$-site cation induces structural modifications, due to an increase or decrease in ionic radii, and changes in the electron count of the partially occupied $B$ site bands, depending on valence. We, therefore, investigate a series of different $A$-site cations, as shown in Fig. 3(a), to identify additional candidates for nickelates which adopt the same crystal structure at ambient pressure as the superconducting phase of $La_3Ni_2O_7$.

The only isoelectronic element that can be substituted on the $A$ site for La is the larger cation, Ac. Indeed $Ac_3Ni_2O_7$ adopts a crystal structure with the $Fmmm$ space group symmetry, as shown in Fig. 3(a), and the band structure, shown in Fig. 3(b), is very similar to that in the high pressure phase of $La_3Ni_2O_7$ shown in Fig. 1(e). Therefore, this material might also be a superconductor at ambient pressure. In practice, however, $Ac_3Ni_2O_7$ is an unlikely contender for real applications due to the strong radioactivity of actinium. We note that phonon calculations shown in Fig. 5 show that this material would be structurally stable.

We also find that substituting other $4f$ elements on the $A$ site does not induce a transition to a crystal structure with $Fmmm$ symmetry. For example, $Ce_3Ni_2O_7$ is electron doped compared to the La compound and the Ce substitution has the effect of increasing $\theta_c$, up to 16°. At the same time, the partially filled $4f$ bands move close to the Fermi level, as shown in Fig. 3(c). This is not unexpected, as the ionic radii of the $4f$ cations decrease with increasing electron count. This finding is also in agreement with DFT+$U$ studies [41,42].

Changing the $A$-site cation from La to a group 2 element provides an alternative path to apply chemical pressure, albeit with the addition now of hole doping the material. Comparing the crystal structure for $Ca_3Ni_2O_7$ to $Sr_3Ni_2O_7$ and finally $Ba_3Ni_2O_7$ we find that both $\theta_c$ and $\theta_{ab}$ are reduced with increased ionic radii, with $Ba_3Ni_2O_7$ exhibiting a structure with the desired $Fmmm$ space group [Fig. 3(d)]. The reduction in valence of the group 2 elements compared to La also acts to hole dope the material, as can be seen in Figs. 3(d) and 3(e), where both $Ba_3Ni_2O_7$ and $Sr_3Ni_2O_7$ have Ni $t_{2g}$-derived bands ($d_{xz}$, $d_{yz}$, and $d_{xy}$) at the Fermi level, in addition to the $e_g$ bands ($d_{x^2-y^2}$ and $d_{z^2}$). Ba substitution is, therefore, an alternate method to stabilize the desired structure at ambient pressure; however, we note that $Ba_3Ni_2O_7$ has not, to the best of our knowledge, ever been synthesized. This may be due to the challenge of obtaining a 4+ oxidation state for the Ni. Additionally, our phonon calculations of this system (see Appendix B) show imaginary phonon bands, suggesting that the structure is not stable. Nevertheless, partial Ba doping could still be a viable route to reduce the octahedral tilts in $La_3Ni_2O_7$.

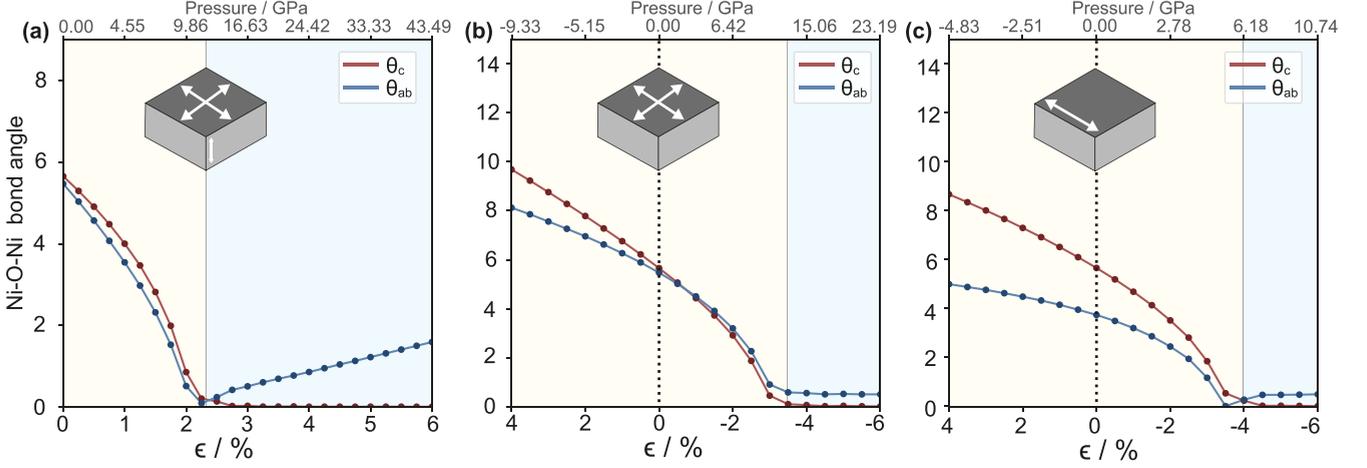

FIG. 4. Comparison of the Ni-O-Ni bond angle for (a) hydrostatic pressure, (b) biaxial strain, and (c) uniaxial strain along the [010] or $b$ axis. Obtained from structural relaxations using DFT+$U$ with $U = 3$ eV applied to the Ni site. These are equivalent calculations to the $U = 0$ eV calculations performed in Fig. 1(c), Fig. 2(a), and Fig. 2(c) of the main text, respectively. The pressure axis was determined from the trace of the stress tensor obtained for the relaxed calculations.

## IV. DISCUSSION

From our work, we have identified multiple strategies to stabilize bilayer nickelates with a crystal structure with the same $Fmmm$ space group as the superconducting high-pressure phase of $La_3Ni_2O_7$. Our study of the influence of strain on the crystal structure shows that uniaxial compression along the $b$ axis as well as biaxial compression can drive a structural transition to a crystal structure with the same space group as the high-pressure phase of $La_3Ni_2O_7$. While the levels of strain are higher than those typically achieved in uniaxial strain tuning [45], it is expected that they are upper limits, as DFT typically underestimates the impact of volume change on the electronic states. Epitaxial strain provides an alternative route to strain MBE-grown thin films of $La_3Ni_2O_7$. MBE growth of the Ruddlesden-Popper series of the lanthanum nickelates has been previously reported [46] with compressive epitaxial strain through growth on $LaAlO_3$ and strain levels of 1–2 % [47]. We also identify chemical means to stabilize a crystal structure with $Fmmm$ space group at ambient pressure through cation substitution on the $A$ site. Through isoelectronic substitution, $Ac_3Ni_2O_7$ adopts a crystal structure with the correct space group, yet is made from radioactive elements. $Ba_3Ni_2O_7$ would also have the correct crystal structure; however, Ba substitution results in significant hole doping and may not be chemically stable. Ba doping, however, would be an interesting candidate to understand how the change in Fermi surface affects the superconductivity of this $Fmmm$ Ni perovskite.

Finally, it is worth highlighting that the structural transition in $La_3Ni_2O_7$ is captured without the inclusion of a Hubbard $U$ term. Although electron correlations are always important in these correlated materials [28], this finding suggests that regardless of the origin or superconductivity the structural transition does not depend on correlations beyond the level of DFT. This raises questions about how such a strong superconducting state can be so sensitive to such a subtle structural transition, suggesting that both electronic and phononic degrees of freedom are important for a full understanding of the superconductivity. Strong coupling phenomena induced by phonon degrees of freedom have recently been identified in infinite layer manganate and nickelate perovskites, where the metal to insulator transition can be explained by elastic fluctuations [48]. The parallel with superconductivity in $La_3Ni_2O_7$ is a topic that will require future investigation.

TABLE I. Lattice parameters, including lattice constants ($a$, $b$, $c$) and out-of-plane and in-plane Ni-O-Ni bond angles ($\theta_c$, $\theta_{ab}$) of the DFT relaxed crystal structures $A_3Ni_2O_7$ discussed in Fig. 3 of the main text. The structural parameters of equivalent calculations performed using DFT+$U$ calculations with $U = 3$ eV on the Ni site are also shown.

| A-site cation | $a$ (Å) | $b$ (Å) | $c$ (Å) | $\theta_c$ (deg) | $\theta_{ab}$ (deg) |
|---|---|---|---|---|---|
| La ($U = 0$ eV) | 5.4155 | 5.5230 | 20.241 | 8.10 | 5.78 |
| La ($U = 3$ eV) | 5.3659 | 5.4303 | 20.052 | 5.66 | 5.47 |
| Ac ($U = 0$ eV) | 5.5205 | 5.5197 | 21.1816 | 0.20 | 0.98 |
| Ac ($U = 3$ eV) | 5.4706 | 5.4706 | 21.4429 | 0.00 | 1.39 |
| Ba ($U = 0$ eV) | 5.6810 | 5.6812 | 20.8785 | 0.00 | 0.26 |
| Ba ($U = 3$ eV) | 5.6590 | 5.6590 | 20.9841 | 0.00 | 0.96 |
| Sr ($U = 0$ eV) | 5.6810 | 5.6812 | 20.8785 | 7.04 | 4.10 |
| Sr ($U = 3$ eV) | 5.4264 | 5.3820 | 20.0062 | 0.00 | 3.92 |
| Ca ($U = 0$ eV) | 5.2357 | 5.4177 | 19.0643 | 14.38 | 11.85 |
| Ca ($U = 3$ eV) | 5.2211 | 5.3820 | 19.0547 | 13.42 | 11.32 |
| Ce ($U = 0$ eV) | 5.2551 | 5.5622 | 20.6211 | 16.68 | 17.70 |
| Ce ($U = 3$ eV) | 5.4530 | 5.6465 | 19.6811 | 18.12 | 15.36 |

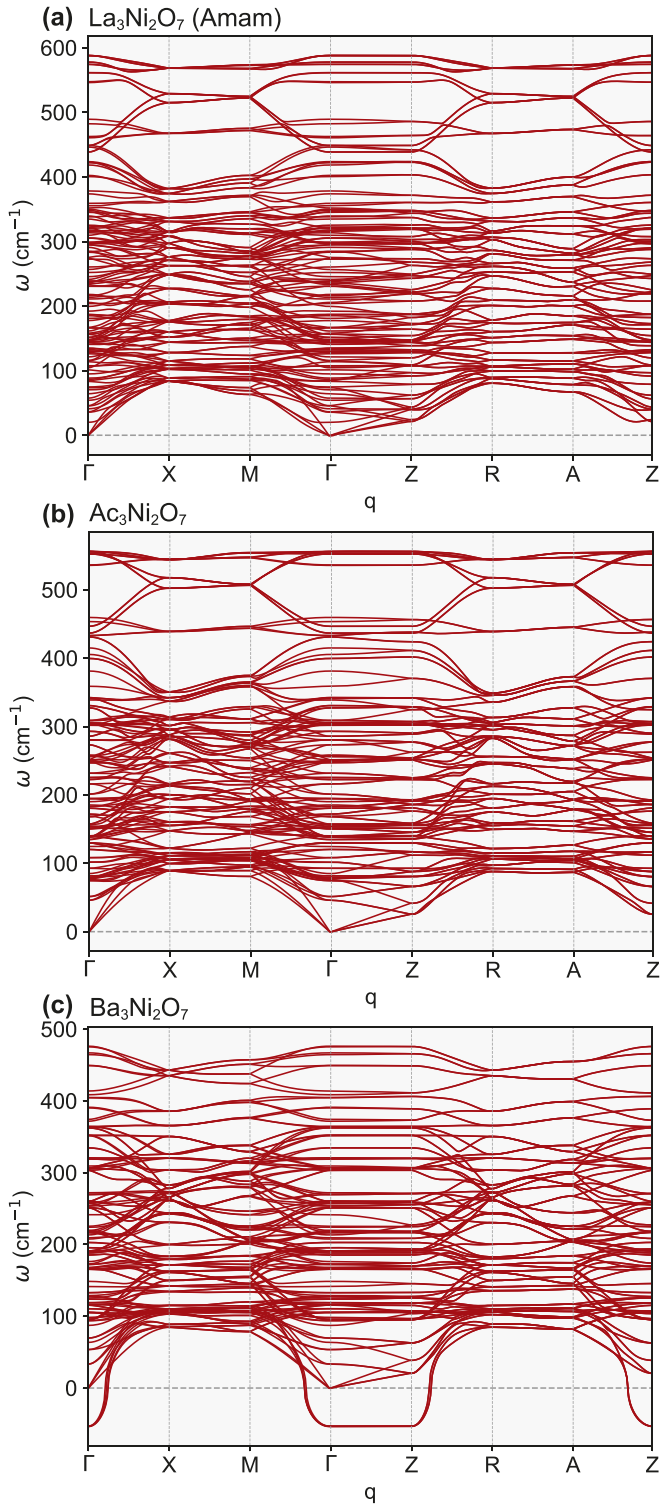

FIG. 5. Calculated phonon dispersion for the DFT relaxed crystals of (a) $La_3Ni_2O_7$, (b) $Ac_3Ni_2O_7$, and (c) $Ba_3Ni_2O_7$.

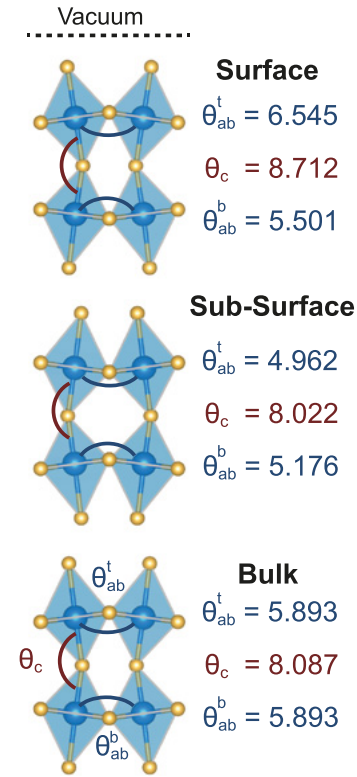

FIG. 6. Structural modifications of the octahedral rotation angles for the surface and subsurface layer of $La_3Ni_2O_7$. Due to the lack of inversion symmetry, there is now a distinction between the upper and lower Ni-O-Ni bond parallel to the *ab* plane. These are defined as $\theta_{ab}^t$ and $\theta_{ab}^b$ for the top and bottom Ni-O-Ni layers, respectively.

*Note added.* Recently, new x-ray diffraction measurements have been reported [49] identifying the superconducting structure as having a space group of $I4/mmm$ rather than $Fmmm$. This is characterized by both $\theta_{ab} = 0$ and $\theta_c = 0$ and a tetragonal, rather than orthorhombic, crystal structure. $Ac_3Ni_2O_7$ and $Ba_3Ni_2O_7$ both relax to this $I4/mmm$ structure as does $La_3Ni_2O_7$ under pressure when applied directly as stress to the energy density functional [42]. The main conclusion of this manuscript—that reducing the *b*-axis lattice constant straightens the bond between octahedra along the *c* direction and induces a transition to a structure with $\theta_c = 0$—remains.

## ACKNOWLEDGMENTS

We thank S.-W. Cheong, A. Gibbs, P. King, P. Littlewood, and A. Rost for insightful discussions. This work used computational resources of the Cirrus UK National Tier-2 HPC Service at EPCC [50] funded by the University of Edinburgh and EPSRC (Grant No. EP/P020267/1). We gratefully acknowledge support from the Leverhulme Trust through Grant No. RPG-2022-315.

## APPENDIX A: INFLUENCE OF HUBBARD $U$ TERM

It is interesting to note that the structural evolution between the *Amam* and *Fmmm* lattices is captured within the framework of DFT, using the PBE exchange-correlation functional, without the inclusion of corrections to account for electron correlations. However, there is growing evidence that a better quantitative agreement between theory and experiment can be obtained if a Hubbard $U$ term of 3 eV is included within the framework of DFT+$U$ to the Ni site [1,42,44]. In Fig. 4 we show repeated structural relaxations for hydrostatic pressure, biaxial strain, and [010] uniaxial strain with the inclusion of a Hubbard $U$ of 3 eV. Qualitatively we find that the trends discussed in the main text remain robust; however, the inclusion of a Hubbard $U$ term reduces the amount of strain or pressure required to induce the transition to the *Fmmm* structure.

Additionally, in Table I we present the relaxed lattice constants and Ni-O-Ni bond angles for the different $A$-site cation discussed in Fig. 3 both for $U = 0$ eV and $U = 3$ eV applied to the Ni site.

## APPENDIX B: PHONON DISPERSION

In order to assess the structural stability of the proposed materials, in Fig. 5 we present phonon-band calculations for $La_3Ni_2O_7$, $Ac_3Ni_2O_7$, and $Ba_3Ni_2O_7$. Phonon calculations were performed using the finite-displacement method and PHONOPY postprocessing software [51,52]. As in Fig. 3, these structures have been fully relaxed, including both internal and lattice degrees of freedom. As shown in Fig. 5, $La_3Ni_2O_7$ and $Ac_3Ni_2O_7$ [Figs. 5(a) and 5(b)] do not exhibit any negative, or imaginary, phonon modes implying that these structures are stable. However, $Ba_3Ni_2O_7$ [Fig. 5(c)] does exhibit an imaginary phonon mode along the $\Gamma$-$Z$ path, which suggests that this structure is not stable.

## APPENDIX C: MODIFICATION AT SURFACES

In the Ruddlesden Popper series of the ruthenates, it is known that the octahedral rotation and tilts can be different at surfaces compared to the bulk [33,34] and induce significant changes in the electronic structure or even the ground states [53,54]. To see whether this is true in $La_3Ni_2O_7$, we have additionally performed structural relaxations on a three-layer thick slab of $La_3Ni_2O_7$ with 15 Å of vacuum on either side of the slab; we then allowed the top two layers to relax, while fixing the bottom layer to the calculated bulk structure. The results are presented in Fig. 6. Surprisingly, we find that the change to both $\theta_c$ and $\theta_{ab}$ is comparatively small, with less than 1° modification compared to bulk for both the surface and subsurface layer. This is in contrast to other perovskites, such as $Sr_2RuO_4$, where an approximately 8° octahedral rotation is induced at the surface [33]. Interestingly, $\theta_c$ increases at the surface, in the opposite direction required to induce the *Fmmm* structure.